\documentclass{SciPost}
\usepackage{amsmath, amssymb}
\usepackage{graphicx, nicefrac, textcomp}
\usepackage{color}
\usepackage{dcolumn}
\usepackage[utf8]{inputenc}
\usepackage{float}
\usepackage{dsfont}
\usepackage{bbold}
\usepackage{svg}
\usepackage[normalem]{ulem}
\usepackage[version=4]{mhchem}
\usepackage[toc,page]{appendix}
\usepackage{changes}
\usepackage{comment}
\usepackage{placeins}
\usepackage{tcolorbox}
\usepackage{tabularx}
\usepackage{etoolbox}
\usepackage{fancyvrb}
\usepackage[shortlabels]{enumitem}
\usepackage{minted}
\usepackage{comment}
\usepackage{seqsplit}
\usepackage{longtable}
\usepackage{tabularx}
\usepackage{xltabular}
\usepackage{subcaption}
\usepackage{bbm}
\usepackage{layouts}
\hypersetup{
    colorlinks,
    linkcolor={red!50!black},
    citecolor={blue!50!black},
    urlcolor={blue!80!black}
}
\usepackage{xcolor}

\newcommand{\ptype}[1]{{\small\textcolor{black!55}{(\texttt{#1})}}\ }
\newcommand{\pdefault}[1]{\par\vspace{2pt}\textcolor{black}{\textit{Default:} \texttt{#1}}}

\usepackage[bitstream-charter]{mathdesign}
\DeclareSymbolFont{usualmathcal}{OMS}{cmsy}{m}{n}
\DeclareSymbolFontAlphabet{\mathcal}{usualmathcal}

\fancypagestyle{SPstyle}{
\fancyhf{}
\lhead{\colorbox{scipostblue}{\bf \color{white} ~SciPost Physics Codebases }}
\rhead{{\bf \color{scipostdeepblue} ~Submission }}

\fancyfoot[C]{\textbf{\thepage}}
}

\BeforeBeginEnvironment{minted}{
\begin{tcolorbox}[colframe=white!75!blue]
}
\AfterEndEnvironment{minted}{
\end{tcolorbox}
}

\mathchardef\up="0222
\mathchardef\dn="0223
\newcommand\ket[1]{\lvert#1\rangle}

\begin{document}
\pagestyle{SPstyle}

\begin{center}{\Large \color{scipostdeepblue}{\textbf{feedback-grape: A JAX-based Python Package for GPU-Accelerated GRAPE with Feedback}}}\end{center}

\begin{center}
Yousef Elbrolosy\textsuperscript{1,2*},
Adrian Bories\textsuperscript{1,3},
Florian Marquardt\textsuperscript{1,3},
Pavlo Bilous\textsuperscript{1}
\end{center}

\begin{center}
{\bf 1} Max Planck Institute for the Science of Light, Staudtstraße 2, 91058 Erlangen, Germany
\\
{\bf 2} German University in Cairo, New Cairo, 11835, Cairo, Egypt
\\
{\bf 3} Department of Physics, Friedrich-Alexander-Universit\"at Erlangen-N\"urnberg, 91058 Erlangen, Germany
\\
*yousefelbrolosy8@gmail.com
\end{center}

\begin{center}
\today
\end{center}

\section*{Abstract}

Modern quantum technology applications require accurate control of quantum systems, including a feedback loop in the control process. While open-loop control can be performed using well established gradient methods like GRadient Ascent Pulse Engineering (GRAPE), it was until recently unclear how to combine these with feedback from measurements, which are inherently non-differentiable. Recent work has successfully demonstrated how feedback can be incorporated into GRAPE using machine learning, resolving this issue. Following this progress, we have developed an efficient Python package implementing the method of ``feedback GRAPE''. In this hands-on tutorial, we discuss the package in detail based on concrete examples. Built on the GPU-accelerated JAX platform, our package offers a user-friendly interface inspired by the widely-used QuTiP library, including simplified treatment of multi-component systems.

\vspace{\baselineskip}

\noindent\textcolor{white!90!black}{%
\fbox{\parbox{0.975\linewidth}{%
\textcolor{white!40!black}{\begin{tabular}{lr}%
  \begin{minipage}{0.6\textwidth}%
    {\small Copyright attribution to authors. \newline
    This work is a submission to SciPost Physics Codebases. \newline
    License information to appear upon publication. \newline
    Publication information to appear upon publication.}
  \end{minipage} & \begin{minipage}{0.4\textwidth}
    {\small Received Date \newline Accepted Date \newline Published Date}%
  \end{minipage}
\end{tabular}}
}}
}

\tableofcontents

\section{Introduction}

The extensively developing field of quantum technologies poses ever higher demands on quantum control techniques \cite{glaser2015training, koch2016controlling}. In absence of a feedback loop and if a high-quality numerical model of the system dynamics is available, well established gradient methods offer a reliable solution. A paradigmatic example of such open-loop quantum control is the method of GRadient Ascent Pulse Engineering (GRAPE) introduced in 2005 in the context of NMR spectroscopy \cite{Khaneja_GRAPE_NMR_2005, machnes2011comparing}. In GRAPE, the gradients of a quality measure (usually based on fidelity) with respect to the controlling fields are evaluated. The latter are then tuned towards maximization of the quality measure. These steps are repeated until convergence is achieved.

If feedback is present, however, the overall control process is not differentiable anymore and GRAPE in its standard form is not applicable. One way to circumvent this problem is offered by model-free reinforcement learning \cite{SuttonBarto, Marquardt_Bukov_2601.18953_2026} as demonstrated in Ref. \cite{fosel2018reinforcement, porotti2022deep, Sivak_PhysRevX.12.011059} and experimentally in Ref. \cite{reuer2023realizing, baum2021experimental}. In this approach, no numerical model of the system dynamics is required and the control tuning is performed based directly on the outcomes of the measurements during and/or at the end of the control process. However, this is disadvantageous in case a good quality model is present, since the discarded knowledge could otherwise be used to improve the search for the optimal control parameters. In particular, the system dynamics between the measurements remains deterministic and differentiable, and should be efficiently incorporated in the control process.

Such a combination of GRAPE and feedback has been achieved recently in Ref. \cite{Porotti_fgrape_PRXQuantum_2023}, where some concepts from model-based reinforcement learning \cite{SuttonBarto, Marquardt_Bukov_2601.18953_2026} were applied to include feedback, whilst performing gradient ascent through the quantum dynamics. Note that in Ref. \cite{Porotti_fgrape_PRXQuantum_2023}, the measurements are parametrized and optimized alongside the control fields. The efficacy of the hereby developed method, ``feedback GRAPE'', was demonstrated on a number of concrete control scenarios \cite{Porotti_fgrape_PRXQuantum_2023, fgrape-GKP} covering state preparation with feedback, state purification and state stabilization with extension to quantum error correction. Given the high potential value to the community, we saw the need to provide an efficient implementation of the methods developed in Ref.~\cite{Porotti_fgrape_PRXQuantum_2023}. Two implementations of feedback-GRAPE are already available online: the code accompanying Ref.~\cite{Porotti_fgrape_PRXQuantum_2023}~\cite{porotti_feedback_grape_github}, and the code used for Gottesman-Kitaev-Preskill (GKP) quantum error correction in Ref.~\cite{fgrape-GKP}~\cite{puviani_gqf_github}. However, both were written for the specific problems studied in those works. Our goal here is to provide a general implementation that can be applied to a wide range of feedback-control tasks.

In this work, we demonstrate an efficient GPU-accelerated Python package primarily designed to perform feedback GRAPE. As the backbone of our codes, we employ the JAX library \cite{jax2018github} which allows to combine physics modeling with machine learning techniques in highly efficient vectorized GPU-accelerated functions. At the same time, we provide a user-friendly interface by following and extending the patterns of the well known QuTiP library \cite{Johansson2013, Nation2014}. To perform feedback GRAPE, the user has to encode the system's Hamiltonian, dissipation channels (if present), a sequence of parametrized control and measurements gates, and the control quality measure, i.e., the reward. Then, by calling a corresponding function from the package, an automated optimization of the control and measurement parameters is triggered. In this hands-on tutorial, we describe these functionalities in detail and demonstrate them using ample code examples. We assume that the user is familiar with general scientific programming in Python using NumPy, and explain the functionalities from other third-party libraries as they are used in the considered examples. The package documentation can be found on the website \cite{fgrape_docs}. The source code is openly available on GitHub \cite{fgrape_github} under the MIT license, and contributions are welcome.

The article is structured as follows. In Section~\ref{sec:install}, we describe the package dependencies and installation process. In Section~\ref{sec:fgrape_algo}, we introduce the feedback GRAPE algorithm. Following that, we show in Section~\ref{sec:qm_components} how basic quantum-mechanical components (e.g. quantum states and operators) are encoded using our package. In Section~\ref{sec:parametrized_cm_gated}, we discuss how the gates parametrization is represented and turned into quantum-mechanical operators in the code. In Section~\ref{sec:example}, we describe two application examples of feedback GRAPE using our package. In Section~\ref{sec:extra_features}, we present further useful features of the package. Finally, we conclude with an outlook and conclusions in Sections~\ref{sec:outlook} and~\ref{sec:conslusions}, respectively.

\section{Dependencies and installation\label{sec:install}}

Table~\ref{tab:dependencies} summarizes the minimal required versions of Python and the third-party Python libraries which are used by our package.
%
%
\begin{table}[h!]
\centering
\begin{tabular}{ccccccccc}
Python & NumPy & SciPy & JAX & Optax & FLAX & Dynamiqs \\\hline
$\geq$ 3.11 & $\geq$ 2.2.4 & $\geq$ 1.15.2 & $\geq$ 0.5.2 & $\geq$ 0.2.4 & $\geq$ 0.10.6 & $\geq$ 0.3.2
\end{tabular}
\caption{Software dependencies and their version requirements. We note that specific versions of \texttt{orbax-checkpoint<=0.11.32} and \texttt{chex>=0.1.89} are also included in the \texttt{requirements.txt} for compatibility.}
\label{tab:dependencies}
\end{table}
Our code employs NumPy \cite{NumPy}, SciPy \cite{scipy}, and Matplotlib \cite{matplotlib}, which are well-established libraries in the scientific community. The backbone of the package is the JAX library \cite{jax2018github}, recently developed at Google. JAX's main functionality consists of highly efficient, GPU-accelerated manipulation of data arrays, represented similarly to NumPy arrays. Additionally, we use the JAX-based packages Optax \cite{jax_ecosys} for optimization and Flax \cite{flax2020github} for defining neural networks. For modeling quantum dynamics, we use our own implementations, except for solving the Lindblad master equation, for which we employ the recently developed Dynamiqs library \cite{dynamiqs2025}, which offers GPU-accelerated and differentiable JAX-based solvers.


The feedback GRAPE Python package can be installed using the \verb+PyPI+ package manager. The version without GPU support is installed as
\begin{verbatim}
    pip install feedback-grape
\end{verbatim}
If an NVIDIA GPU is present on the machine, it can be leveraged by the package to significantly accelerate the computations. The GPU-capable version is installed as
\begin{verbatim}
    pip install feedback-grape[cuda12]
\end{verbatim}
which assumes compatibility with CUDA 12. Running one of these commands without further parameters will automatically identify and install/update the dependencies. Note that the two variations differ by triggering the installation of different JAX versions: without and with GPU support, respectively. As an alternative to installing via the \verb+PyPI+ manager, the code can be cloned directly from the GitHub repository \cite{fgrape_github}.

\section{Feedback-GRAPE algorithm\label{sec:fgrape_algo}}

\begin{figure}[H]
        \centering
        \includegraphics[width=\linewidth]{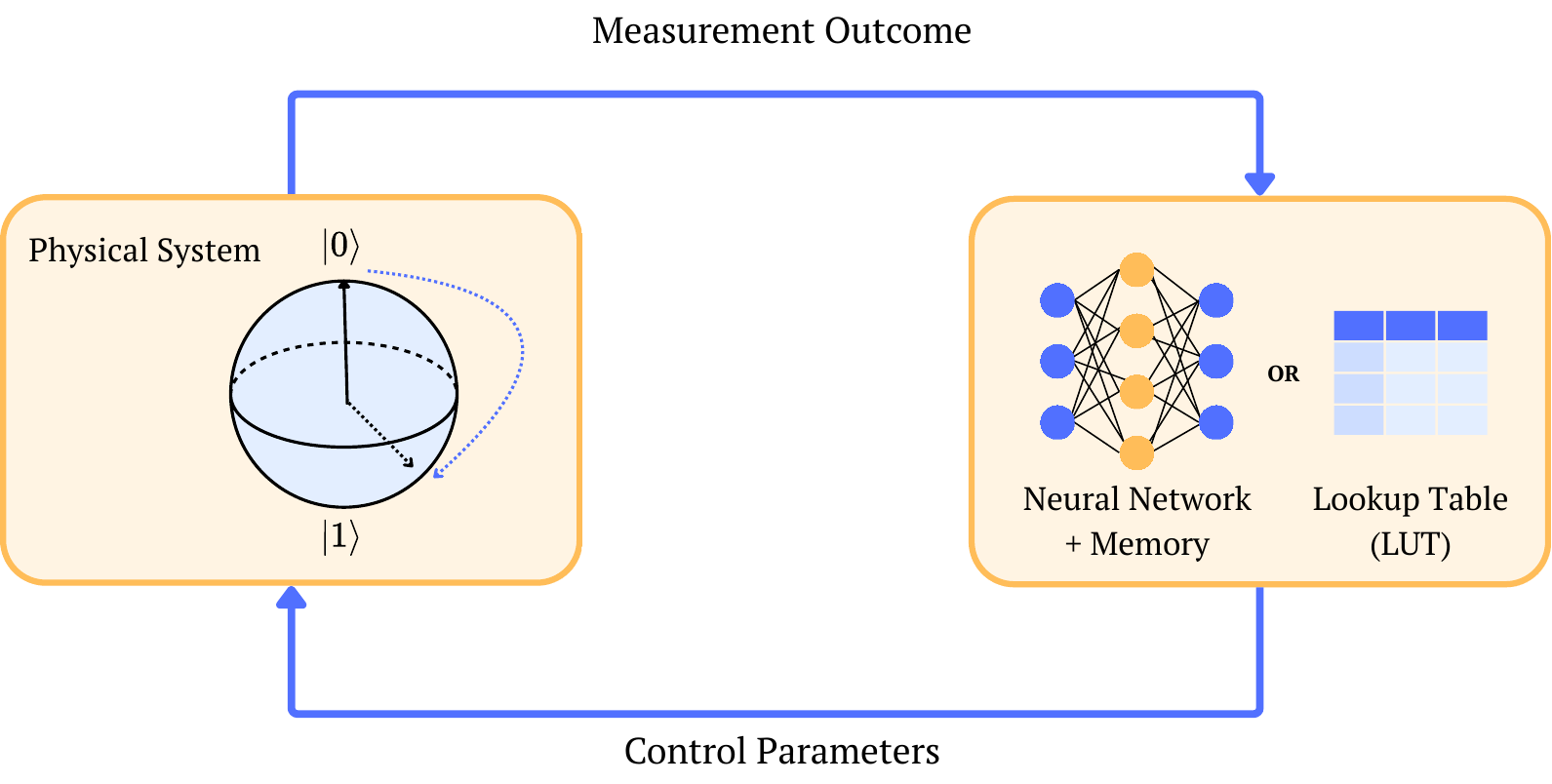}
        \caption{A high-level overview of the feedback-GRAPE algorithm. A physical system is controlled using a feedback loop mechanism: Measurement outcomes from the quantum system are fed into a recurrent neural network or a lookup table which generates the control parameters for the next time step.}
        \label{fig:rl-diagram}
\end{figure}

In this section, we present a high-level overview of the feedback GRAPE algorithm sketched in Fig.~\ref{fig:rl-diagram} and introduced in Ref. \cite{Porotti_fgrape_PRXQuantum_2023}. A typical control scenario is represented by a circuit consisting of multiple repeated ``elementary'' control blocks. In the following, these repetitions are referred to as ``timesteps''. Each timestep contains unitary control and measurement gates. Both gate types are parametrized and optimized in feedback GRAPE.

The measurement gates extend the ordinary GRAPE algorithm and play a two-fold role. Firstly, they constitute non-unitary control elements removing  entropy from the system and purifying its state. Secondly, they yield information on which the parameters of the subsequent gates are conditioned. In the present incarnation of our package, described in the following, we consider generalized binary measurements and denote their outcomes $\pm 1$. Each measurement is therefore represented by an operator pair $\hat {M}_{\pm 1}$ satisfying the completeness condition for positive operator-valued measures (POVMs) $\hat {M}_{+1}^\dagger\hat{M}_{+1}^{\vphantom{\dagger}} + \hat{M}_{-1}^\dagger \hat{M}_{-1}^{\vphantom{\dagger}} = \mathbbm{1}$. If a system in a state described by the density matrix $\rho$ is measured, the probability $P_x$ of a particular outcome $x=\pm 1$ and the corresponding post-measurement state $\rho_x$ can be obtained as
\begin{equation}
P_x = \mathrm{Tr}\left( \rho \hat{M}_x^\dagger \hat{M}_x\right)\;,\qquad\rho_x = \frac{\hat{M}_x \rho \hat{M}_x^\dagger}{P_x}\;.
\end{equation}

The stochastic measurement outcomes are used to choose the control and measurement parameters in the future steps using a feedback policy. The latter is either a lookup table (LUT) containing directly the optimized parameters or a recurrent neural network (RNN). We note that both approaches maintain memory over the past measurement outcomes in the current trajectory, thus rendering the feedback non-Markovian. The described approach closely follows reinforcement learning \cite{SuttonBarto}. However, exploration here isn't a design choice, but is forced by the stochasticity of the measurements.

Training is performed by collecting statistics over multiple trajectories and evaluating the average reward gradients over the trainable parameters using the equation in Ref. \cite{Porotti_fgrape_PRXQuantum_2023}:
\begin{equation}\label{eq:reward-update}
    \frac{\partial \left\langle R(m) \right\rangle}{\partial \theta}
    =
    \left\langle
    \frac{\partial R(m)}{\partial \theta}
    \right\rangle
    +
    \left\langle
    R(m)\frac{\partial \ln P_\theta(m)}{\partial \theta}
    \right\rangle\;.
\end{equation}
Here, the trainable parameters are denoted as $\theta$. The averaging is performed over a batch of trajectories each identified by its sequence of measurement outcomes denoted as $m$. For each such sequence $m$, we denote $P(m)$ the trajectory probability and $R(m)$ the reward gained in it. The reward choice depends on the control task and will be discussed in the sections devoted to corresponding examples. The obtained gradients are then used to perform gradient ascent towards the optimal control parameters. Further details will be discussed in the context of the concrete examples introduced in the following sections. For more details on the feedback-GRAPE algorithm and equation \ref{eq:reward-update}, the reader is referred to Ref. \cite{Porotti_fgrape_PRXQuantum_2023}.

\section{Encoding a quantum-mechanical system\label{sec:qm_components}}

We begin now with demonstration of our feedback GRAPE Python package. Since the user has to encode the physical system by hand, we start here from the basic quantum mechanical entities (quantum states, operators etc.) intended for this purpose. These are implemented in the \verb+feedback_grape.utils+ subpackage demonstrated in the following.

As the first example, we create a qubit state $(1, 0)^\dagger$ and a cavity coherent state $\ket{\alpha}$ with $\alpha=3$:
\begin{minted}[breaklines]{python}
from feedback_grape.utils.states import basis, coherent

N_cav = 20
basis_state = basis(n=2)
coherent_state = coherent(n=N_cav, alpha=3)
\end{minted}
\noindent Here, we first imported the corresponding functions \verb+basis+ and \verb+coherent+ from the submodule \verb+feedback_grape.utils.states+, and called them with the corresponding arguments. The parameter $N_\mathrm{cav}$ denotes the dimension of the truncated Hilbert space of the cavity. Note that the shown code closely follows the QuTiP syntax. The important difference to QuTiP is that the invoked functions return not objects of custom classes, but usual JAX arrays. These can be manipulated using the standard JAX techniques, which are similar to those for NumPy arrays, see e.g. the JAX documentation \cite{jax2018github}. For instance, a density matrix can be constructed from a pure state using directly the JAX means. An alternative way is to use the function \verb+feedback_grape.utils.fidelity.ket2dm+ in the same way as in QuTiP.

Next, we demonstrate the most elementary quantum operators by constructing the number operator from the bosonic creation and annihilation operators:
\begin{minted}[breaklines]{python}
from feedback_grape.utils.operators import create, destroy

number_operator = create(N_cav) @ destroy(N_cav)
\end{minted}
\noindent As in the case of states, operators are represented by JAX arrays. Consequently, and in contrast to QuTiP, the operator multiplication is performed via the $@$ operation, whereas the $*$ operation means element-wise multiplication of two matrices. Other operators and operator-related functions defined in \verb|feedback_grape.utils.operators| can be found in Table \ref{tab:operators} and will be used throughout this article.

%
%
\begin{table}[h]
\centering
\begin{tabularx}{\linewidth}{lX|lX}
\hline
\textbf{Function} & \textbf{Description} & \textbf{Function} & \textbf{Description} \\
\hline
\verb|sigmax| & Pauli $\hat \sigma_x$ operator & \verb|create| & Bosonic creation operator \\
\verb|sigmay| & Pauli $\hat\sigma_y$ operator & \verb|destroy| & Bosonic annihilation operator \\
\verb|sigmaz| & Pauli $\hat\sigma_z$ operator & \verb|identity| & Identity operator \\
\verb|sigmap| & Pauli $\hat\sigma_+$ operator & \verb|cosm| & Cosine of an operator \\
\verb|sigmam| & Pauli $\hat\sigma_-$ operator & \verb|sinm| & Sine of an operator \\
\hline
\end{tabularx}
\caption{Operators and functions defined in \texttt{utils.operators}.}
\label{tab:operators}
\end{table}

As the next step, we show an important function \verb+tensor+, used for creating states and operators for compound systems. Here, we construct a system consisting of a cavity coupled to a qubit. In this example, the ground state of this system is created with the following code:

\begin{minted}[breaklines]{python}
from feedback_grape.utils.tensor import tensor

ground_qubit = basis(n=2)
ground_cavity = basis(n=N_cav)
ground_state = tensor(ground_cavity, ground_qubit)
\end{minted}
\noindent The \verb+tensor+ function can also be applied to quantum operators represented by 2D JAX arrays. Note that whereas a system with few components can be encoded using such \verb+tensor+-ing, this approach becomes cumbersome for more components, as e.g. qubit chains. For this case, we have implemented an alternative convenient interface described in Section~\ref{sec:composite_sys_interface}.

Two other important functions of our package are \verb+fidelity+ and \verb+purity+ in the submodule \verb+feedback_grape.utils.fidelity+. For two quantum states characterized by density matrices $\rho$ and $\sigma$, the \verb+fidelity+ function computes the Uhlmann-Jozsa fidelity \cite{Jozsa_UhlmannFidelity} given by the formula $F(\rho, \sigma) = \left(\text{tr}\sqrt {\sqrt{\rho}\sigma\sqrt{\rho}}\right)^2$. The \verb+purity+ function evaluates the quantum state purity $P(\rho) = \text{tr}\left(\rho^2\right)$. Depending on the control task, \verb+fidelity+ and \verb+purity+ can be used to construct the trajectory reward.

This finishes the description of the basic components for encoding quantum systems using our package. As an alternative to the described functionalities, the users can create basic quantum-mechanical entities directly in QuTiP and then apply the \verb|jaxify| function from the \verb|feedback_grape.utils.operators| submodule in order to obtain a corresponding JAX representation, see Section~\ref{sec:from_qutip}.

\section{Parametrized control and measurement gates\label{sec:parametrized_cm_gated}}

The feedback-GRAPE method aims to achieve the control goal by optimizing the parameters of the control and measurement gates. In this section, we discuss how these can be encoded in our package.

\subsection{Description of a control example\label{sec:control_desc}}

\begin{figure}
        \centering
        \includegraphics[width=\linewidth]{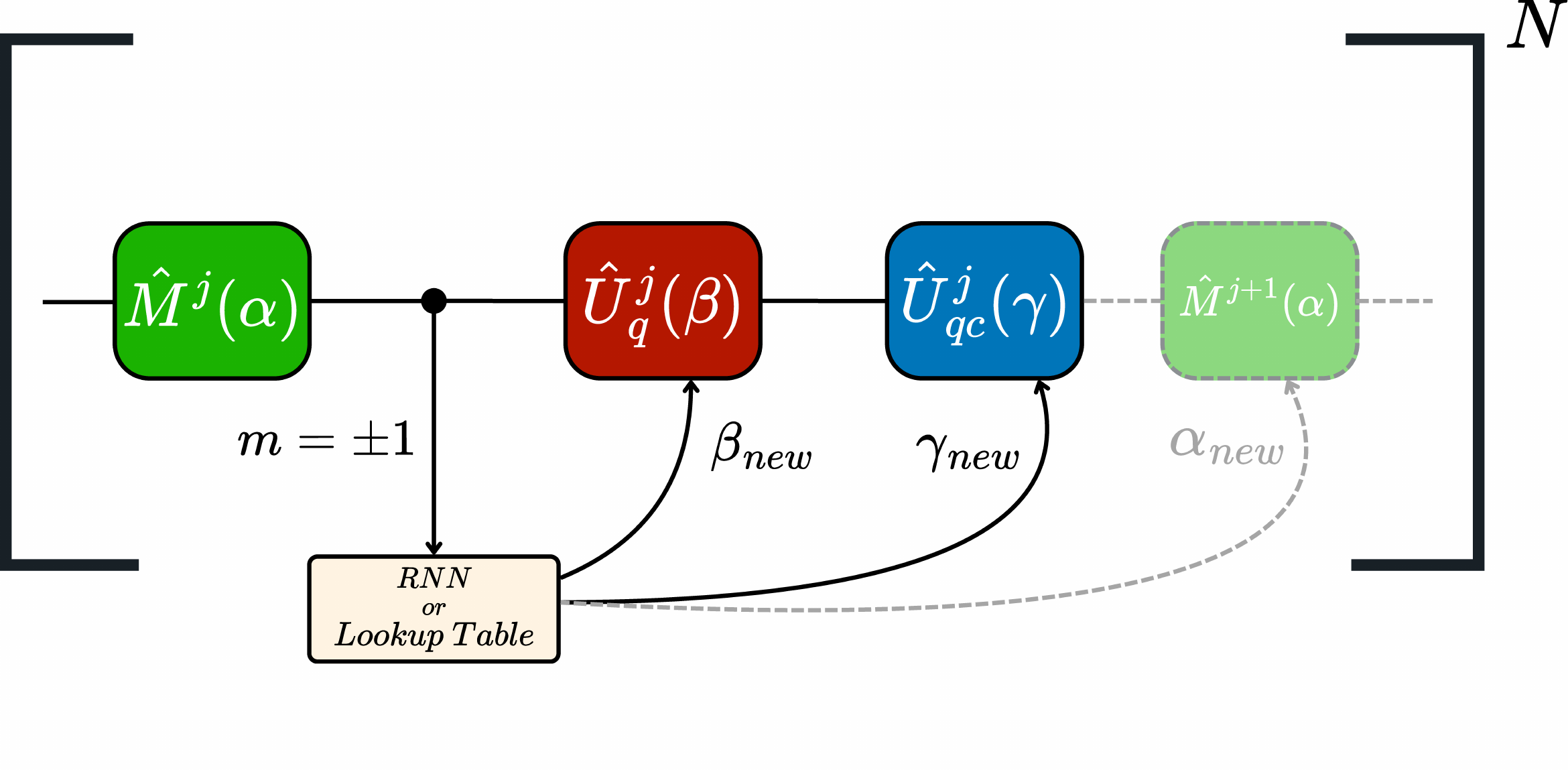}
        \caption{Control loop of a qubit-cavity system consisting of $N$ timesteps. Measurement outcomes are fed to a LUT or RNN, which predicts the parameters for the subsequent operations, including the next measurement (dashed). In this work, only binary measurements are considered.}
        \label{fig:loop-diagram}
\end{figure}

We now turn to a concrete example of a cavity coupled to two qubits, following Example C in Ref.~\cite{Porotti_fgrape_PRXQuantum_2023}: a control qubit, which is driven and interacts with the cavity, and an ancilla qubit, which is used only to measure the cavity. This system is subject to a control sequence illustrated in Fig.~\ref{fig:loop-diagram}. Each timestep is repeated $N$ times and consists of a parametrized measurement $\hat M$ followed by two parametrized unitary gates $\hat U_q$ and $\hat U_{qc}$ described below. The measurement yields one of two possible outcomes, which is fed to a LUT or RNN. The latter predicts the gate parameters for the following operations, including the measurement in the next timestep (dashed).

In this example, the measurement is represented by the following operator pair
\begin{eqnarray}
\hat M_{+1}(\gamma, \delta) &=& \cos(\gamma \hat a^\dagger \hat a + \delta / 2)\;,\label{eq:M_plus1}\\
\hat M_{-1}(\gamma, \delta) &=& \sin(\gamma \hat a^\dagger \hat a + \delta / 2)\;,\label{eq:M_minus1}
\end{eqnarray}
where $\hat a$ is the photon annihilation operator, and $\gamma$ and $\delta$ are the (real) parameters to be optimized. Physically, this measurement is performed by (i) putting the ancilla qubit in a state lying on the equator of the Bloch sphere; (ii) letting it freely evolve under the coupling to the cavity; (iii) projectively measuring the ancilla qubit in a basis in the equatorial plane, which yields a binary outcome. The parameter $\gamma$ includes the ancilla qubit-cavity interaction strength and the evolution time in (ii), whereas $\delta$ represents the angle between the axes in (i) and (iii).

The parametrized unitary gates are given by the following equations:
\begin{eqnarray}
    \hat U_q(\alpha) &=& \exp[-i(\alpha \hat\sigma_+ + \alpha^* \hat\sigma_-) / 2]\;,\label{eq:U_q}\\
    \hat U_{qc}(\beta) &=& \exp[-i(\beta \hat a \hat\sigma_+ + \beta^* \hat a^\dagger \hat\sigma_-) / 2]\;.\label{eq:U_qc}
\end{eqnarray}
The gate $\hat U_q(\alpha)$ represents an external drive on the control qubit with $\alpha$ containing the drive strength and time. The gate $\hat U_{qc}(\beta)$ describes the interaction between the control qubit and the cavity with the coupling strength and the interaction time contained in the parameter $\beta$. Note that $\alpha$ and $\beta$ in Eqs. (\ref{eq:U_q}, \ref{eq:U_qc}) are complex, and the actual optimized parameters are their real and imaginary parts.

\subsection{Unitary operations and measurements as Python functions\label{sec:op_as_pyfunc}}

In order to represent the behavior of a parametrized control (measurement) gate in the code, the user creates a function receiving the parameters (and the measurement outcome) as arguments and returning the corresponding concrete matrix. Below, we will employ the functionalities described in Section \ref{sec:qm_components}, which are imported as:
\begin{minted}[breaklines]{python}
import jax
import jax.numpy as jnp
from jax.scipy.linalg import expm
from feedback_grape.utils.operators import (
    sigmap, sigmam, create, destroy, identity, cosm, sinm
)
\end{minted}
\noindent We now encode the measurement given by Eqs. (\ref{eq:M_plus1}, \ref{eq:M_minus1}):
\begin{minted}[breaklines]{python}
def meas_operator(measurement_outcome, params):

    gamma, delta = params
    cav_operator = (
        gamma * create(N_cav) @ destroy(N_cav)
        + delta / 2  * identity(N_cav)
    )
    angle = tensor(cav_operator, identity(2))
    meas_op = jnp.where(
        measurement_outcome == 1,
        cosm(angle),
        sinm(angle),
    )
    
    return meas_op
\end{minted}
\noindent The optimization routine calling this function will assume that the measurement outcome is passed as the first argument, whereas the optimized parameters are provided in an array as the second argument. Note that it is important to use here the matrix \verb+cosm+ / \verb+sinm+ functions, and not \verb+jnp.cos+ / \verb+jnp.sin+ acting on each matrix element independently. Additionally, note that the conditional statement in the code is represented using the \verb|jnp.where| function rather than the standard Python \verb|if-else| clause. This is necessary for compatibility of the created function with JAX's just-in-time (JIT) compilation \cite{jax2018github}, which significantly improves the computation speed and is crucial for many functionalities of the feedback GRAPE package.

The parametrized unitary control gates are represented in the code as follows:
\begin{minted}[breaklines]{python}
def qubit_unitary(alphas):

    alpha_re, alpha_im = alphas
    alpha = alpha_re + 1j * alpha_im
    h_dt = (alpha*sigmap() + alpha.conjugate()*sigmam()) / 2
    
    return tensor(identity(N_cav), expm(-1j * h_dt))
\end{minted}
\begin{minted}[breaklines]{python}
def qubit_cavity_unitary(betas):

    beta_re, beta_im = betas
    beta = beta_re + 1j * beta_im
    a_sigp = tensor(destroy(N_cav), sigmap())
    adag_sigm = tensor(create(N_cav), sigmam())
    h_dt = (beta * a_sigp + beta.conjugate() * adag_sigm) / 2
    
    return expm(-1j * h_dt)
\end{minted}
\noindent Here, the optimization parameters are passed to the functions as a single array. We stress again that both $\alpha$ and $\beta$ in Eqs. (\ref{eq:U_q}, \ref{eq:U_qc}) are generally speaking complex. Therefore, each function above receives an array of two real arguments, which are the real and the imaginary part of the corresponding complex parameter.

\subsection{Python class for managing optimized parameters\label{sec:gate-class}}

As exemplified above, the users encode the behavioral pattern of parametrized control and measurement gates as Python functions receiving the optimization parameters as arguments. In this section we present a Python class \verb+Gate+ which stores the actual parameter values and manages the concrete action of the operation (unitary control or measurement) on the physical system. Objects of the \verb+Gate+ class are created from the following three main components: (i) a function \verb+gate+ encoding the operation as described in Section \ref{sec:op_as_pyfunc}; (ii) an array \verb+initial_params+ with the initial parameter values (usually random); (iii) a boolean flag \verb+measurement_flag+ indicating if the operation is a measurement. There is also another optional argument \texttt{param\_constraints} (not used here) which serves as a constraint for the optimized parameters, see Ref.~\cite{fgrape_docs} for details. The parameters can be initialized using the JAX randomization mechanism, which we briefly describe in the paragraph below. The objects of the \verb+Gate+ class for the measurement and the unitary gates defined in Section~\ref{sec:op_as_pyfunc} are instantiated as follows:
\begin{minted}[breaklines]{python}
from feedback_grape.fgrape import Gate

key = jax.random.key(seed=52)
key, subkey = jax.random.split(key)

measure = Gate(
    gate=meas_operator,
    initial_params=jax.random.uniform(
        subkey,
        shape=(2,), # gamma, delta
        minval=-jnp.pi,
        maxval= jnp.pi,
    ),
    measurement_flag=True,
)
\end{minted}

\begin{minted}[breaklines]{python}
key, subkey = jax.random.split(key)

qub_unitary = Gate(
    gate=qubit_unitary,
    initial_params=jax.random.uniform(
        subkey,
        shape=(2,), # alpha_re, alpha_im
        minval=-jnp.pi,
        maxval= jnp.pi,
    ),
    measurement_flag=False,
)
\end{minted}
\begin{minted}[breaklines]{python}
key, subkey = jax.random.split(key)

qub_cav = Gate(
    gate=qubit_cavity_unitary,
    initial_params=jax.random.uniform(
        subkey,
        shape=(2,), # beta_re, beta_im
        minval=-jnp.pi,
        maxval= jnp.pi,
    ),
    measurement_flag=False,
)
\end{minted}

\noindent Since the seed determines the initial parameters, and optimization outcomes can be sensitive to this initialization, we advise users to similarly explore multiple seeds.

\paragraph{Randomization mechanism in JAX.} The JAX randomization strategy \cite{jax2018github} significantly differs from the more common approach based on a hidden randomizer state (like e.g. in NumPy). The latter would be incompatible with JIT compilation which assumes that the compiled function be ``pure'', i.e. with an outcome depending only on the function arguments and without any hidden state influencing it. Therefore, the randomizer state in JAX is encapsulated in a so called ``random key'' which is passed explicitly to a function performing randomization, e.g. \verb+jax.random.uniform+ or \verb+jax.random.normal+. The first key is created from a seed, whereas new random keys are created by ``splitting'' the old ones. Using the same key twice in a particular randomizing function leads to the same output\footnote{Note, however, that cases have been reported \cite{jax_nondeterm_discussion}, in which the same random key leads to different outcomes on different platforms, so this ``deterministic'' behavior should not be relied on.}. Moreover, reusing a key in different functions can still lead to correlated results and is therefore considered a bad practice. It is common to split a key into two keys just before invoking a function with randomization. One of the keys is used as the called function argument, whereas the other one is kept for later randomization using the described approach. Note that by convention these two keys are often assigned to the variable names  \verb+subkey+ and \verb+key+, respectively. We refer the user to the JAX documentation \cite{jax2018github} for more details and examples on the JAX randomization mechanism.

\section{Illustration examples\label{sec:example}}

At the moment, our package addresses state preparation, purification and stabilization tasks, with fidelity and purity as the available reward functions. In this section, we showcase our code on two concrete examples of closed-loop quantum control: state preparation from a thermal initial state and state stabilization. As the feedback policy, we use LUTs and RNNs, respectively.

\subsection{Example task: Cavity state preparation with feedback}

Following the description of the control example in Section \ref{sec:control_desc} and Example C in Ref.\cite{Porotti_fgrape_PRXQuantum_2023}, we illustrate preparation of the following superposition of the Fock state in the cavity:
\begin{eqnarray}
\ket{\psi_\text{target}} = \frac{1}{\sqrt3}(\ket{1} + \ket{2} + \ket{3})\;.
\end{eqnarray}
We assume here that the system is initially in a thermal state, which necessitates measurements in the preparation procedure. The initial cavity state reads
\begin{equation}
\rho_0 = \frac{1}{Z} \sum_{n=0}^{N_{cav} - 1} e^{-\beta n} \, |n\rangle \langle n|\;, \qquad \beta = \ln\left(\frac{1}{\bar{n}} + 1\right)\;,
\end{equation}
where for this example the average number of photons $\bar{n}=1$ is assumed.

Using the utilities introduced in Section \ref{sec:qm_components}, the density matrices representing the initial and the target states are encoded as:
\begin{minted}[breaklines]{python}
n_average = 1
beta = jnp.log((1 / n_average) + 1)
diags = jnp.exp(-beta * jnp.arange(N_cav))
normalized_diags = diags / jnp.sum(diags, axis=0)
rho_cav0 = jnp.diag(normalized_diags)
rho_qub0 = basis(2, 0) @ basis(2, 0).conj().T
rho0 = tensor(rho_cav0, rho_qub0)
\end{minted}

\noindent For the target state, we use the \verb|fock| function (an alias for \verb|basis|) to construct Fock states of the cavity:

\begin{minted}[breaklines]{python}
from feedback_grape.utils.states import fock

psi_cav_target = (
    fock(N_cav, 1) + fock(N_cav, 2) + fock(N_cav, 3)
) / jnp.sqrt(3)
psi_qub_target = basis(2, 0)
psi_target = tensor(psi_cav_target, psi_qub_target)
rho_target = psi_target @ psi_target.conj().T
\end{minted}

As the next step, we arrange the \verb+Gate+ objects created here and in Section~\ref{sec:parametrized_cm_gated} in a Python list. The order in the list corresponds to the application order of the corresponding operations in a timestep (which in this example will be repeated 5 times):
\begin{minted}[breaklines]{python}
system_params = [measure, qub_unitary, qub_cav]
\end{minted}

The control procedure is now defined and we can switch to the search of the optimal gate parameters. This is achieved by calling the function \verb+optimize_pulse+, which is the central tool in our package. We show below how it can be done for our particular example, and summarize more generally the \verb+optimize_pulse+ arguments in Table~\ref{table:optimize-pulse-parameters}. We also show the type of each variable following the standard type annotation rules in Python.
\begin{minted}[breaklines]{python}
from feedback_grape.fgrape import optimize_pulse

result = optimize_pulse(
    init_state=rho0,
    target_state=rho_target,
    system_params=system_params,
    num_time_steps=5,
    mode="lookup",
    goal="fidelity",
    max_iter=3000,
    learning_rate=0.02,
    evo_type="density",
    progress=True,
    batch_size=10,
    eval_time_steps=5
)
\end{minted}

\newcolumntype{L}[1]{>{\raggedright\arraybackslash}p{#1}}
\begin{xltabular}{\textwidth}{p{4.5cm}>{\raggedright\arraybackslash}X}
\caption{
Summary of arguments of the \texttt{optimize\_pulse} function with their types shown following standard Python type annotation rules.
}
\label{table:optimize-pulse-parameters} \\
\hline
\textbf{Parameter} & \textbf{Description} \\
\hline
\endfirsthead
\multicolumn{2}{c}{\tablename\ \thetable\ -- \textit{Continued}} \\
\hline
\textbf{Parameter} & \textbf{Description} \\
\hline
\endhead
\hline
\multicolumn{2}{r}{\textit{Continued on next page}} \\
\endfoot
\hline
\endlastfoot
\texttt{init\_state} & \ptype{jnp.ndarray}Initial state or density matrix. \\ \\
\texttt{target\_state} & \ptype{jnp.ndarray}Target state or density matrix. \\ \\
\texttt{system\_params} & \ptype{list[Gate]}List of \texttt{Gate} objects in the order of the corresponding operations in a timestep. \\ \\
\texttt{num\_time\_steps} & \ptype{int}Number of repeated timesteps in a trajectory. \\ \\
\texttt{max\_iter} & \ptype{int}Maximum number of optimization iterations. \\ \\
\texttt{convergence\_threshold} & \ptype{float | None}The smallest improvement in mean reward, measured over \texttt{convergence\_window} iterations. Provide \texttt{None} to disable convergence checking and always run for \texttt{max\_iter} iterations. \\ \\
\texttt{convergence\_window} & \ptype{int}Number of iterations averaged into each of the two disjoint windows compared when checking convergence. Only used when \texttt{\seqsplit{convergence\_threshold}} is not \texttt{None}.\pdefault{50} \\ \\
\texttt{convergence\_patience} & \ptype{int}Number of consecutive window comparisons that must fall below \texttt{convergence\_threshold} before stopping. Only used when \texttt{convergence\_threshold} is not \texttt{None}.\pdefault{3} \\ \\
\texttt{learning\_rate} & \ptype{float}Learning rate used by the Adam optimizer \cite{kingma2014adam}, employed for gate parameter search. \\ \\
\texttt{evo\_type} & \ptype{str}Type of state representation used during evolution, e.g., \texttt{"state"} or \texttt{"density"}. \\ \\
\texttt{reward\_weights} & \ptype{list[float | int] | None}Weights for the reward at each time step. If \texttt{None}, only the final step is weighted. \\ \\
\texttt{goal} & \ptype{str}Optimization objective: \texttt{"purity"}, \texttt{"fidelity"}, or \texttt{"both"}.\pdefault{\texttt{"fidelity"}} \\ \\
\texttt{batch\_size} & \ptype{int}Number of trajectories evolved in parallel during training.\pdefault{1} \\ \\
\texttt{eval\_batch\_size} & \ptype{int}Number of trajectories used during evaluation.\pdefault{10} \\ \\
\texttt{eval\_time\_steps} & \ptype{int | None}Number of time steps used during evaluation. If \texttt{None}, defaults to \texttt{num\_time\_steps}. \\ \\
\texttt{mode} & \ptype{str}Feedback policy: \texttt{"nn"} (neural network) or \texttt{"lookup"} (lookup table).\pdefault{\texttt{"lookup"}} \\ \\
\texttt{rnn} & \ptype{callable}Recurrent model used when \texttt{mode = "nn"}. Ignored otherwise. \\ \\
\texttt{rnn\_hidden\_size} & \ptype{int}Hidden dimension of the RNN model. Only used with \texttt{mode = "nn"}.\pdefault{30} \\ \\
\texttt{progress} & \ptype{bool}If \texttt{True}, prints progress (cost every 10 iterations). Useful for debugging but may slow optimization.\pdefault{\texttt{False}} \\ \\
\texttt{\seqsplit{initial\_trainable\_parameters}} & \ptype{jnp.ndarray}Parameters to warm start from, continuing training from a previous \texttt{FgResult}'s \texttt{\seqsplit{optimized\_trainable\_parameters}} instead of a fresh initialization.\pdefault{\texttt{None}} \\
\hline
\end{xltabular}

After the optimization process is complete, an object of a custom class \verb|FgResult| is returned. It has a number of attributes intended for representing the computation results and their further use. Those attributes are discussed in Table \ref{table:fg-result}.

\begin{xltabular}{\textwidth}{l >{\raggedright\arraybackslash}X}
\caption{Summary of fields in \texttt{FgResult}.}
\label{table:fg-result} \\
\hline
\textbf{Field} & \textbf{Description} \\
\hline
\endfirsthead
\multicolumn{2}{c}{\tablename\ \thetable\ -- \textit{Continued}} \\
\hline
\textbf{Field} & \textbf{Description} \\
\hline
\endhead
\hline
\multicolumn{2}{r}{\textit{Continued on next page}} \\
\endfoot
\hline
\endlastfoot
\texttt{\seqsplit{optimized\_trainable\_parameters}} & \ptype{jnp.ndarray}Optimized control amplitudes. \\ \\
\texttt{iterations} & \ptype{int}Number of iterations taken for optimization. \\ \\
\texttt{final\_state} & \ptype{jnp.ndarray}Final operator after applying the optimized control amplitudes. \\ \\
\texttt{returned\_params} & \ptype{list[jnp.ndarray]}Array of final parameters for each time step. \\ \\
\texttt{final\_purity} & \ptype{jnp.ndarray | None}Final purity of the optimized control. \ptype{None}when \texttt{goal} does not include purity. \\ \\
\texttt{final\_fidelity} & \ptype{jnp.ndarray | None}Final fidelity of the optimized control.  \ptype{None}when \texttt{goal} does not include fidelity. \\ \\
\texttt{fidelity\_each\_timestep} & \ptype{list[jnp.ndarray] | None}Fidelity values of the optimized control at each time step and batch.  \ptype{None}when \texttt{goal} does not include fidelity. \\ \\
\texttt{purity\_each\_timestep} & \ptype{list[jnp.ndarray] | None}Purity values of the optimized control at each time step and batch.  \ptype{None}when \texttt{goal} does not include purity. \\ \\
\texttt{reward\_history} & \ptype{jnp.ndarray} Reward at each training iteration.\\
\hline
\end{xltabular}

\subsubsection*{Numerical results}

\begin{figure}[H]
    \centering
    \includegraphics[width=0.85\linewidth]{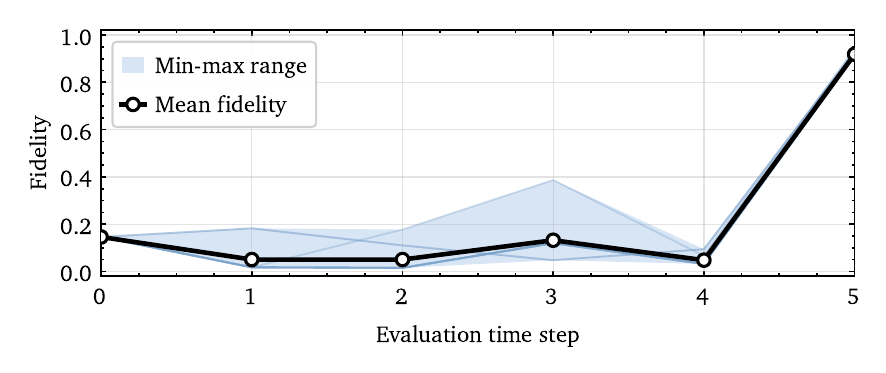}
    \caption{Fidelity evolution during evaluation of the trained        
  feedback-GRAPE procedure, where only the fidelity at the final time step was optimized. Faint blue curves show the individual    
  stochastic measurement trajectories under the learned look-up policy.
   The shaded band shows their min–max range, and the solid black curve their
   mean. Note that some trajectories coincide and are indistinguishable in the plot.
   }
    \label{fig:fidelity-evolution}
\end{figure}

The training in this example is performed for 3000 iterations. After the optimization process is complete, we extract the evaluation fidelity in each time step via \texttt{\seqsplit{result.fidelity\_each\_timestep}} and examine how the state evolves with time.
As shown in Fig.~\ref{fig:fidelity-evolution}, the parameters converged to a state which optimizes the fidelity in the final time step (in this example, \(t = 5\)). This behavior is natural for the state preparation task. Note that other tasks, e.g. state stabilization, may require keeping the fidelity high over many time steps. This can be controlled by the argument \texttt{reward\_weights} of the \texttt{optimize\_pulse} function which we discuss and demonstrate in the next example.

\begin{figure}[H]
    \centering
    \includegraphics[width=\linewidth]{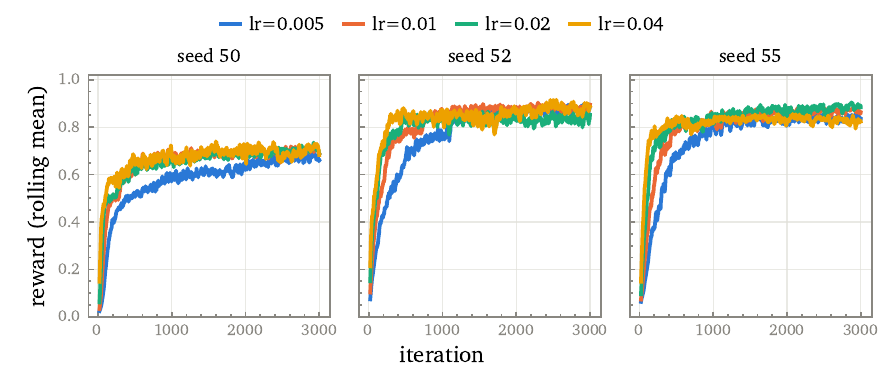}
    \caption{Progress with training iterations for different random seeds and learning rates (indicated as ``lr''). The training was performed in all cases for 3000 iterations.
    %
   }
    \label{fig:seeds-results}
\end{figure}

The progress in the training process can be investigated using the \texttt{\seqsplit{result.reward\_history}} attribute. The optimization procedure may be sensitive to the parameter initialization and the hyperparameter values, which should be explored by the users for the case at hand. In Fig. \ref{fig:seeds-results} we show the result of such exploration for the present example. In each case, the training was performed for 3000 iterations. It can be seen that the convergence pattern and the result indeed shows dependence on the initial random seed and the learning rate (indicated as ``lr''). We attribute this behavior to complex reward landscape with many local minima.

We note that we ran the training for a fixed number of iterations and do not employ the early stopping mechanism. The latter is available in our package and can be leveraged when calling the \texttt{optimize\_pulse} function by setting the following arguments (see also Table \ref{table:optimize-pulse-parameters}):
\begin{itemize}
    \item \texttt{\seqsplit{convergence\_threshold}} -- The smallest improvement in mean reward, measured over \texttt{convergence\_window} iterations.
    \item \texttt{\seqsplit{convergence\_window}} -- Number of iterations averaged into each of the two disjoint windows compared when checking convergence.
    \item \texttt{\seqsplit{convergence\_patience}} -- Number of consecutive window comparisons that must fall below \texttt{convergence\_threshold} before stopping.  
\end{itemize}


%

\subsection{Example task: State stabilization}
\label{sec:state_stabilization}

In this section we consider the task of stabilization of a quantum state in presence of decoherence. We depart from the system considered in the previous section and showcase stabilization with the feedback-GRAPE method for a system of five qubits subject to amplitude damping and dephasing described below. The code for this example is available in the GitHub repository~\cite{stabil_5qub_github}; since it closely follows the structure of the state-preparation task, we highlight here only the aspects specific to this example. Specifically, we aim stabilize the following superposition state:
\begin{equation}
\ket{\psi} =  \frac{1}{\sqrt{2}}(\ket{00000} + \ket{11111})\;.
\end{equation}
In general, the evolution of the density matrix $\rho$ in presence of noise can be modeled using the Lindblad equation:
\begin{eqnarray}\label{lindblad-eq}
    \frac{d\rho}{dt} = -\frac{i}{\hbar} [\hat{H}, \rho] + \sum_k \left( \hat c_k \rho \hat c_k^\dagger - \frac{1}{2} \{\hat c_k^\dagger \hat c_k, \rho\} \right)\;,
\end{eqnarray}
where $\hat H$ is the Hamiltonian governing the unitary (noiseless) evolution and $\hat c_k$ are the jump operators of the noise channels. We assume that the system idles for time $t_\text{idle}$ with $\hat H = 0$ and consider each qubit to undergo amplitude damping and dephasing modeled by the jump operators $\hat c_- = \sqrt{\kappa_-} \hat{\sigma}_-$ and $\hat c_z = \sqrt{\kappa_z}\hat{\sigma}_z$ with the rates $\kappa_-$ and $\kappa_z$. Under these assumptions, the decoherence result is fully determined by the products $\gamma = \kappa t_\text{idle}$. In this example we assume $\gamma_- = \gamma_z = 0.1$.

We encode this by passing the jump operators of the system into a new class called \texttt{Decay} as follows:
\begin{minted}[breaklines]{python}
from feedback_grape.utils.operators import (
    sigmaz, sigmam, identity
)
gamma_z = 0.1 # dephasing rate
gamma_m = 0.1 # decay rate
I = identity(2)
\end{minted}
\begin{minted}[breaklines]{python}
dephasing = [
    gamma_z**0.5 * tensor(sigmaz(), I, I, I, I),
    gamma_z**0.5 * tensor(I, sigmaz(), I, I, I),
    gamma_z**0.5 * tensor(I, I, sigmaz(), I, I),
    gamma_z**0.5 * tensor(I, I, I, sigmaz(), I),
    gamma_z**0.5 * tensor(I, I, I, I, sigmaz()),
]

decay = [
    gamma_m**0.5 * tensor(sigmam(), I, I, I, I),
    gamma_m**0.5 * tensor(I, sigmam(), I, I, I),
    gamma_m**0.5 * tensor(I, I, sigmam(), I, I),
    gamma_m**0.5 * tensor(I, I, I, sigmam(), I),
    gamma_m**0.5 * tensor(I, I, I, I, sigmam()),
]

decay_gate = Decay(
    c_ops=dephasing + decay,
)
\end{minted}

\noindent The \verb|Decay| object takes \texttt{c\_ops} as input, which is a list of JAX arrays corresponding to the jump operators\footnote{For an alternative way to define quantum operators for compound systems without explicit tensoring, see Section~\ref{sec:composite_sys_interface}}. This \texttt{\seqsplit{decay\_gate}} object is then inserted in the \texttt{system\_params} list where the measurement and control gates are accumulated to form the structure of a time step. 
When the \texttt{\seqsplit{decay\_gate}} is applied, the Lindblad equation~\eqref{lindblad-eq} with $\hat{H} = 0$ is integrated numerically over a unit time interval using the master-equation solver of the \texttt{dynamiqs} library~\cite{dynamiqs2025}.

In our time-step ansatz, the decay gate is followed by a sequence of four fully parametrized generalized measurements interleaved with fully parametrized unitaries. We note that such encoding of decay as a discrete gate can be used to approximately model decoherence during the control. This is achieved by repeatedly interrupting the control process and inserting partial decay gates; see Ref.~\cite{Porotti_fgrape_PRXQuantum_2023} for details.

For the state preparation task considered in Section \ref{sec:example}, the reward was naturally determined by the fidelity in the last time step. For stabilization, it is important to incorporate the fidelity also in the intermediate time steps. In our package, this can be achieved using the \verb+reward_weights+ argument of the \verb+optimize_pulse+ function. For $N$ time steps, it receives an array with weights $[w_1, \ldots, w_N]$. Then the reward is expressed as $R = \sum_{i=1}^N w_i F_i$, where $F_i$ is the fidelity in the $i$-th time step. The default value is $[0, 0, \ldots, 0, 1]$, which was implicitly used in the state preparation example.

\begin{figure}[H]
    \centering
    \includegraphics[width=\linewidth]{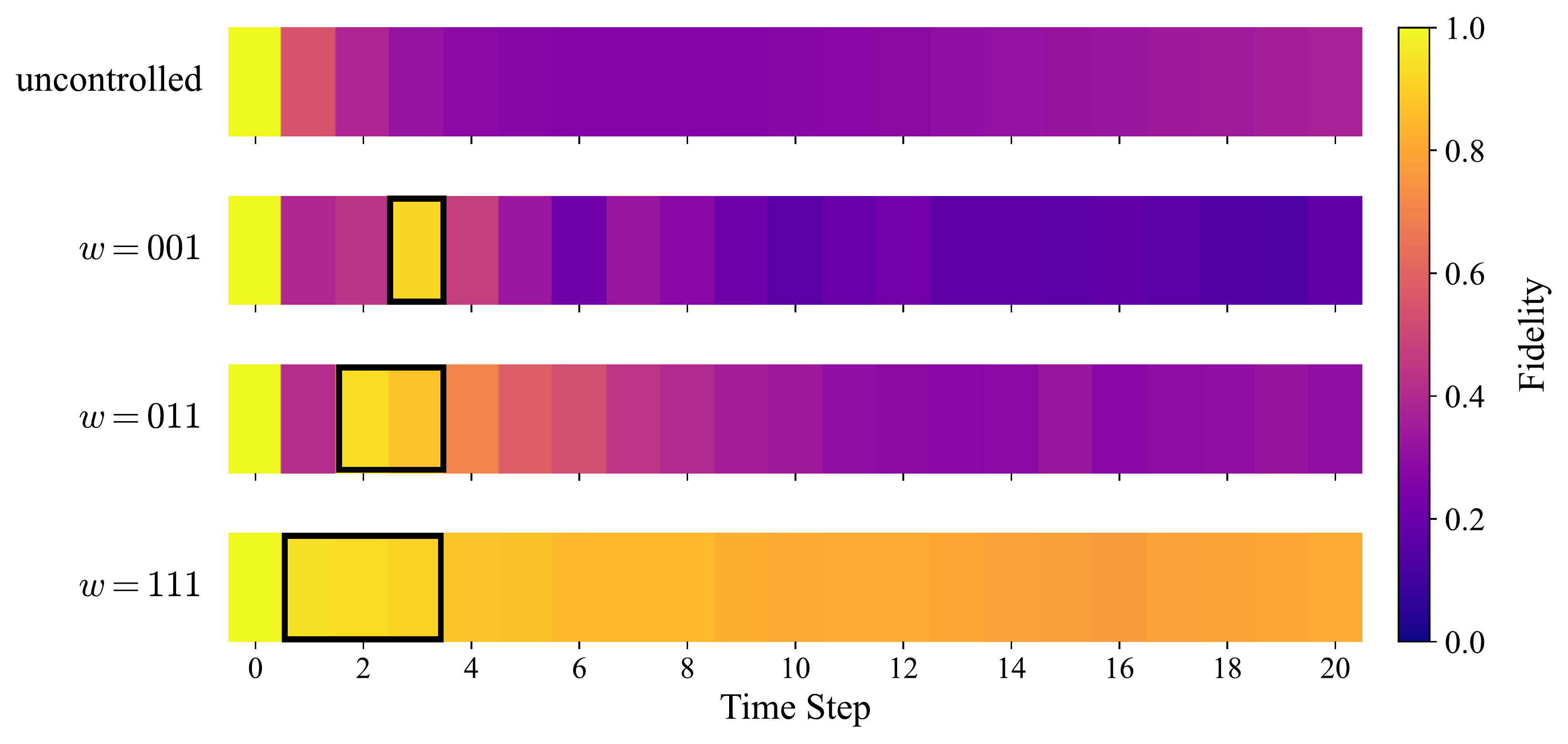}
    \caption{Comparison of ensemble fidelity of RNN models trained for state stabilization of a five-qubit superposition state $\ket{\psi} = \frac{1}{\sqrt{2}}(\ket{00000} + \ket{11111})$ against amplitude damping and dephasing. The training was performed for three distinct values passed to the \texttt{reward\_weights} argument: (from top to bottom) $[0,0,1]$, $[0,1,1]$ and $[1,1,1]$. The time-steps whose fidelity enters the reward are highlighted by the bold black contour and the fidelities at the 0th time step represent the initial state before any gates are applied. The results are compared with an uncontrolled system (top row).
    }
    \label{fig:reward-weights}
\end{figure}

In Fig.~\ref{fig:reward-weights} we show the fidelity achieved in each time step of the stabilization process, using a neural network for the optimization (\verb+mode="nn"+ in \verb+optimize_pulse+), compared against an uncontrolled system. The RNN was trained on 3 time steps and applied to stabilize the state over 20 time steps. For each set of training parameters, we trained three randomly initialized RNNs and selected the best performing model based on its average fidelity over 20 time steps. We demonstrate the results for three distinct values passed to the \verb+reward_weights+ argument: $[0,0,1]$, $[0,1,1]$ and $[1,1,1]$. The time-steps whose fidelity enters the reward are highlighted by the bold black contour. We see that the stabilization is efficient only if the fidelities in all three time-steps contribute to the reward $R$.

Using the exemplary control tasks for state preparation from a thermal state and state stabilization of a state subject to decoherence, we have covered the most relevant functionalities of our package. Below, we show other helpful features and refer the user to the package documentation \cite{fgrape_docs}  for further details.

\section{Discussion of other features\label{sec:extra_features}}

In this section we introduce further features helpful for solving quantum control tasks.


\subsection{GRAPE without feedback}
Our package supports the original GRAPE algorithm \cite{Khaneja_GRAPE_NMR_2005} for optimizing control pulses in absence of feedback. Note that we do not use the analytical approach presented in Ref.~\cite{Khaneja_GRAPE_NMR_2005}, but rather employ automatic differentiation offered by the JAX library.

There are two ways to perform GRAPE optimization within the package. The first approach uses the same gated pipeline described in the preceding sections, but without supplying a measurement operator. In this case, the control sequence consists entirely of parametrized unitary gates, and the optimization proceeds exactly as in the feedback-GRAPE setting, except that no measurement-conditional branching occurs. There, the user passes \texttt{mode="no-measurement"} in the \texttt{\seqsplit{fgrape.optimize\_pulse}} function and proceeds as described in the preceding sections.

The second approach follows the traditional GRAPE formulation more closely. Here, the user directly supplies the system Hamiltonian, the initial and target states, and the number of time intervals into which the control pulse is discretized. This is accessed through the interface demonstrated in the following code snippet:
\begin{minted}[breaklines]{python}
from feedback_grape.grape import optimize_pulse

res = optimize_pulse(
    H_drift=H0,
    H_control=H_ctrl,
    U_0=psi0,
    C_target=psi_target,
    num_t_slots=num_of_intervals,
    total_evo_time=T,
    evo_type=evolution_type,
    optimizer=optimizer_type,
)
\end{minted}
\noindent Note that we imported here from \texttt{\seqsplit{feedback\_grape.grape}} rather than \texttt{\seqsplit{feedback\_grape.fgrape}}.

A comprehensive description of the traditional GRAPE in our package with example notebooks for both methods are available in the package documentation \cite{fgrape_docs}.

\subsection{Composite system interface\label{sec:composite_sys_interface}}

Alongside the normal QuTiP-like API based on explicitly constructing tensor products for subsystem operators, we provide also an alternative API which allows users to avoid manually tensoring susbystem operators with identities and therefore can be convenient for working with large compound systems. This API is modular and can be expanded to a general composite system structure. However it is currently implemented only for systems consisting of qubits and cavities, and we refer to it as to QubitCavity API.

The central object of this API is the \verb|QubitCavity| class defined as follows:
\begin{minted}[breaklines]{python}
class QubitCavity:
    def __init__(self, num_qubits: int, *cavity_dims: int):
        """
    
        Qubit comes first in the Hilbert space, followed by
        cavities.
    
        Args:
            num_qubits (int): Number of qubits.
            *cavity_dims (int): Dimensions of each cavity.
        """
\end{minted}
\noindent Here, the first argument specifies the number of qubits, and the remaining arguments specify the cavity dimensions. For example, a system of two qubits and two cavities can be encoded as follows:
\begin{minted}[breaklines]{python}
from feedback_grape.utils.modeling import QubitCavity
qc = QubitCavity(2, 10, 20)
\end{minted}
\noindent where we model the first 10 Fock states of the first cavity and the first 20 Fock states of the second cavity. The individual subsystems can then be addressed explicitly. For example, the first qubit and the first cavity are accessed as:
\begin{minted}[breaklines]{python}
q0 = qc.qubits[0]
c0 = qc.cavities[0]
q0_sigmaz = q0.sigmaz
c0_annihilation = c0.a
\end{minted}
\noindent Here, \verb+q0.sigmaz+ returns the Pauli operator $\sigma_z$ acting on qubit 0 and the identity elsewhere, and \verb+c0.a+ returns the annihilation operator acting on cavity 0 and the identity elsewhere. This is done automatically without requiring the user to manually tensor with identity operators for the spectator subsystems. More examples of the \verb|QubitCavity| API are given in the documentation \cite{fgrape_docs}.

\subsection{Conversion of operators from QuTiP\label{sec:from_qutip}}

QuTiP is one of the most used platforms for simulating quantum dynamics and it offers an extensive library of predefined operators. However, the corresponding Python objects are incompatible with JAX. Therefore, we implemented a function \verb+jaxify+ converting QuTiP's object to JAX arrays which can be used in our package. For example the \verb|sigmax| operator (which is also directly available in our package) can be converted from QuTiP as follows:
\begin{minted}[breaklines]{python}
import qutip as qt
from feedback_grape.utils.operators import jaxify

sigx = jaxify(qt.sigmax())
\end{minted}




\section{Outlook\label{sec:outlook}}

The current implementation of \texttt{feedback-grape} provides a flexible and efficient framework for feedback-based quantum optimal control. In the following we point out several possible extensions which can be implemented in the future.

Attention-based neural networks such as transformers have revolutionized many fields, for instance, natural language processing, genomics, and computer vision. Recently, the use of transformers for quantum feedback control was demonstrated \cite{transformer-grape}. Such architecture could be added to the package as an extra option for finding the optimized parameters which may help capture long-range temporal correlations more efficiently than the present models.

The interface for compound systems demonstrated in Section~\ref{sec:composite_sys_interface} currently supports only systems consisting of qubits and cavities. Since this API is especially advantageous for systems with many components, its generalization would be helpful for usage of the package in such cases.


In the current version of the package, the optimal control strategy for producing, purifying or stabilizing a specific quantum state is sought. An extension to ensembles of initial states would enable e.g. discovery of protocols for quantum error-correction, which can be treated as stabilization of a subspace of states rather than a single state.

In addition, the library of numerical solvers can be further expanded. Different physical regimes and control scenarios may benefit from specialized integration schemes or propagation methods with improved efficiency, stability, or memory scaling. Providing a wider set of solvers would allow users to better tailor simulations to their specific applications. 

Finally, a further useful extension would be the option to supply a custom reward function. Currently, the optimization target is fixed to fidelity, purity, or a combination thereof; allowing users to define arbitrary reward functions would enable optimization for other quantities relevant to their specific application.



\section{Conclusion\label{sec:conslusions}}

We have developed and demonstrated a general and efficient GPU-accelerated Python
package implementing the method of feedback GRAPE. We illustrated its use
on two concrete closed-loop control tasks --- cavity state preparation
from a thermal state and multi-qubit state stabilization --- and described
its further features, including standard GRAPE without feedback, a
composite-system interface, and conversion of QuTiP operators. Built on
the JAX library, the package provides a user-friendly interface that
follows and extends the patterns of the well-known QuTiP library. We warmly invite contributions from the community and hope that this package helps further advancements in the field of feedback-based quantum optimal control.



\bibliography{refs}

\end{document}